\documentclass[twocolumn,english,american,prl,american,aps,showkeys,showpacs,reprint]{revtex4-1}
\usepackage{lmodern}

\usepackage[T1]{fontenc}
\usepackage[latin9]{inputenc}
\usepackage{xcolor}
\usepackage{pdfcolmk}
\usepackage{babel}
\usepackage{verbatim}
\usepackage{units}
\usepackage{amsmath}
\usepackage{amssymb}
\usepackage{graphicx}
\PassOptionsToPackage{normalem}{ulem}
\usepackage{ulem}
\usepackage[unicode=true,pdfusetitle,
 bookmarks=true,bookmarksnumbered=false,bookmarksopen=false,
 breaklinks=false,pdfborder={0 0 1},backref=false,colorlinks=false]
 {hyperref}

\makeatletter

\providecolor{lyxadded}{rgb}{0,0,1}
\providecolor{lyxdeleted}{rgb}{1,0,0}
\@ifundefined{textcolor}{}
{%
 \definecolor{BLACK}{gray}{0}
 \definecolor{WHITE}{gray}{1}
 \definecolor{RED}{rgb}{1,0,0}
 \definecolor{GREEN}{rgb}{0,1,0}
 \definecolor{BLUE}{rgb}{0,0,1}
 \definecolor{CYAN}{cmyk}{1,0,0,0}
 \definecolor{MAGENTA}{cmyk}{0,1,0,0}
 \definecolor{YELLOW}{cmyk}{0,0,1,0}
 }

\makeatother

\begin{document}

\title{Optics of a Gas of Coherently Spinning Molecules}

\author{Uri Steinitz, Yehiam Prior, and Ilya Sh. Averbukh}

\affiliation{Department of Chemical Physics, Weizmann Institute of Science, 234
Herzl Street, Rehovot 76100, Israel}

\email{ilya.averbukh@weizmann.ac.il}

\selectlanguage{american}%

\begin{abstract}
We consider optical properties of a gas of molecules that are brought to fast unidirectional spinning by a pulsed laser field. It is shown that a circularly polarized probe light passing through the medium inverts its polarization handedness and experiences a frequency shift controllable by the sense and the rate of molecular rotation.  Our analysis is supported by two recent experiments on the laser-induced rotational Doppler effect in molecular gases, and provides a good qualitative and quantitative description of the experimental observations.
\end{abstract}

\date{\today}

\maketitle
When a circularly polarized photon is scattered forward from an anisotropic
body, its polarization handedness may be inverted, and the scattering
is accompanied by an exchange of angular momentum $\Delta L=2\hbar$
between the photon and the body. For a body rotating at
a frequency $\Omega$ much smaller than the light frequency, the kinetic
energy of rotation is modified by $\Omega\Delta L=2\hbar\Omega$. This should be
compensated by the photon energy change, therefore the frequency of the scattered photon
becomes shifted by $2\Omega$. This phenomenon, called the rotational
Doppler shift\cite{Garetz1981a,Bretenaker1990,Allen1994,Bialynicki-Birula1997,Bialynicki-Birula2012}
is a macroscopic classical analog of the rotational Raman effect \cite{Cabannes1929}.
It has been observed in the past by using mechanical rotation of
optical elements \cite{Allen1966,Garetz1979,Courtial1998,Lavery2013}
and electro-optic effects in a nonlinear crystal subject to a rotating
microwave electric field \cite{Buhrer1962}.

In recent years there has been an ever-growing interest in aligning
gas molecules by ultra-short laser pulses (for recent reviews see
\cite{Fleischer2012,Ohshima2010}, earlier developments are described
in \cite{Stapelfeldt2003}). The research resulted in numerous achievements
ranging from differentiation of molecular species through enhancement
of filamentation effects to high harmonic generation control and attosecond pulse generation. Several methods
have been suggested and demonstrated for converting the  transient molecular alignment into a concerted unidirectional
molecular rotation, including the techniques of {}`optical centrifuge'
\cite{Karczmarek1999,Villeneuve2000,Yuan2011}, {}`molecular propeller'
\cite{Fleischer2009,Kitano2009,benzene}, and {}`chiral
train' of laser pulses \cite{Zhdanovich2011,Bloomquist2012,Floss2012}.

In this theoretical paper we investigate the propagation of light through a gas of unidirectionally
rotating molecules, analyze polarization and spectral changes due to the energy and angular momentum exchange with the gas, and demonstrate that
the light may experience a THz-range rotational Doppler frequency shift under experimentally feasible conditions. Our analysis considers in a unified
way two recent experiments on the optics of gases of coherently spinning molecules \cite{Korech2013,Korobenko2013} whose results are in accord with our theory.

Consider an ensemble of unidirectionally rotating linear molecules
prepared by one of the above techniques \cite{Karczmarek1999,Villeneuve2000,Yuan2011,Fleischer2009,Kitano2009,benzene,Zhdanovich2011,Bloomquist2012,Floss2012}.
The simplest of them ({}`molecular propeller' \cite{Fleischer2009,Kitano2009,benzene}) uses a sequential excitation by two delayed pump pulses with different linear polarizations.
The first pulse aligns the molecules and the second one applies a biased torque to them thus causing the unidirectional rotation. We denote the
molecular polarizability anisotropy by $\Delta\alpha=\alpha_{\parallel}-\alpha_{\perp}$
($\alpha_{\parallel},\alpha_{\perp}$ are the polarizability values
along and perpendicular to the molecular axis, respectively). The
pump pulses are assumed to be collinear and to propagate at the same
speed, $c/n$, as in the undisturbed and unaligned medium. Here $c$
is the speed of light, $n=\sqrt{1+N\bar{\alpha}\epsilon_{0}^{-1}}$
is the refractive index, $\bar{\alpha}=\left(\alpha_{\parallel}+2\alpha_{\perp}\right)/3$
is the orientation-averaged molecular polarizability, $N$ the
concentration of the molecules, and $\epsilon_{0}$ is the vacuum permittivity. For undepleted pump pulses (as is
the case for nonresonant excitation in transparent gases), the rotational
dynamics of molecules at any given distance down the propagation line
depends only on the time elapsed since the pulsed pump acted at that
same location. Following the excitation, the molecules continue their
field-free rotation, each at a constant angular velocity.

The electric field $\mathbf{E}_{i}$ of an incident probe pulse follows
Maxwell`s equations
\begin{equation}
\partial_{zz}\mathbf{\mathbf{E}}_{i}-\frac{1}{c^{2}}\partial_{tt}\mathbf{E}_{i}=\frac{1}{c^{2}}\partial_{tt}\left(\overleftrightarrow{\chi}\mathbf{E}_{i}\right)\quad,\label{eq:max basic-1}
\end{equation}
 where $\overleftrightarrow{\chi}$ is the electric susceptibility
tensor of the medium. In the paraxial approximation the only significant
field components are those perpendicular to the propagation axis $z$.
We express the susceptibility in the basis of circular polarization
(CP) states,\foreignlanguage{english}{
\begin{eqnarray}
\overleftrightarrow{\chi}\left(z,t\right) & = & N\left(\overline{\alpha}-\frac{\Delta\alpha}{2}\left(\overline{\cos^{2}\theta_{0}}-\frac{1}{3}\right)\right)I+\label{eq:gyrot_alpha}\\
 & + & \frac{N\Delta\alpha}{2}\begin{pmatrix}G & \rho^{*}\\
\rho & G
\end{pmatrix}\quad;\nonumber \\
\rho & = & \left\langle \sin^{2}\theta e^{2i\varphi}\right\rangle \quad;\nonumber \\
G & = & \left\langle \cos^{2}\theta_{0}-\cos^{2}\theta\right\rangle \quad.\nonumber
\end{eqnarray}
}Here $I$ is the unity matrix, and the averaging is made over all
the molecules in a unit volume. We use the standard spherical angles
$\theta$ and $\varphi$ to denote the instantaneous orientation of
the molecules (the polar angle $\theta$ is measured from the propagation direction $z$).
The angle $\theta_{0}$ denotes $\theta$ at the moment of the probe arrival to the molecule's location.
The molecular rotation frequency is
assumed to be small compared with the optical carrier frequency
$\omega_{i}$, thus  allowing the
calculation of the susceptibility by the coordinate transformation to
the molecules' rotating frame.

In what follows, we consider a delayed probe pulse, $\mathbf{E}_{i}=\overrightarrow{\mathcal{E}_{i}}\left(z,t\right)\exp i\omega_{i}\left(t-\nicefrac{z}{V_{i}}\right)$,
with a slowly varying envelope $\overrightarrow{\mathcal{E}_{i}}\left(z,t\right)$
(of two CP components) propagating collinearly with the pumps. We use the first, time-independent,
term of the polarizability in Eq.\ (\ref{eq:gyrot_alpha}), to define
the velocity value $V_{i}=c\left(1+\epsilon_{0}^{-1}N\left(\overline{\alpha}-\nicefrac{\Delta\alpha}{2}\left(\overline{\cos^{2}\theta_{0}}-\nicefrac{1}{3}\right)\right)\right)^{-\nicefrac{1}{2}}$ that is the common term for both CP components.
We rearrange Eq.\ (\ref{eq:max basic-1}) by neglecting the second
order derivatives, and transforming to the probe's time frame $\left(z',\tau\right)\equiv\left(z,t-\nicefrac{z}{V_{i}}\right)$:

\begin{eqnarray}
\partial_{z'}\overrightarrow{\mathcal{E}_{i}} & = & -i\beta_{i}\begin{pmatrix}G & \rho^{*}\\
\rho & G
\end{pmatrix}\overrightarrow{\mathcal{E}_{i}}+2\frac{\beta_{i}}{\omega_{i}}\partial_{\tau}\left[\begin{pmatrix}G & \rho^{*}\\
\rho & G
\end{pmatrix}\overrightarrow{\mathcal{E}_{i}}\right]\quad;\label{eq:dEdz'}\\
\beta_{i} & = & \frac{\mu_{0}N\Delta\alpha\omega_{i}V_{i}}{4}\quad, \nonumber
\end{eqnarray}
where $\mu_{0}$ is the vacuum permeability. Here, $\rho$ and $G$ (and all other time-varying quantities) depend only
on the time $\tau$, elapsed since the pumps arrived at each location.
This is justified as long as the pump-probe
speed difference is small enough so that the probe {}`surfs' on
the seemingly unchanging wake of the pump throughout the 
interaction length, $l$ (i.\,e., when $l\ll\Delta t_{al}c/(N \left|\Delta \alpha \right| \epsilon_0^{-1} )$,
where $\Delta t_{al}$ is the typical time scale of the molecular
alignment dynamics).

We use a basis transformation $\mathcal{R}$ to diagonalize the matrix
that appears in Eq.\ (\ref{eq:dEdz'})

\begin{eqnarray*}
\begin{pmatrix}G & \rho^{*}\\
\rho & G
\end{pmatrix} & = & \begin{pmatrix}-\frac{\rho^{*}}{\left|\rho\right|} & \frac{\rho^{*}}{\left|\rho\right|}\\
1 & 1
\end{pmatrix}\cdot\begin{pmatrix}G-\left|\rho\right| & 0\\
0 & G+\left|\rho\right|
\end{pmatrix}\cdot\begin{pmatrix}-\frac{\rho}{2\left|\rho\right|} & \frac{1}{2}\\
\frac{\rho}{2\left|\rho\right|} & \frac{1}{2}
\end{pmatrix}\\
 & \equiv & \mathcal{R}\left(\tau\right)\,\cdot\mathcal{K}\left(\tau\right)\,\cdot\mathcal{R}^{-1}\left(\tau\right)\quad.
\end{eqnarray*}
Here, again, $\mathcal{R}$ and $\mathcal{K}$ are functions of $\tau$
only. Physically, the transformation $\mathcal{R}$ defines the instantaneous birefringence axes of the medium. We present the two components of the pulse in the $\mathcal{K}$
eigenvectors basis as $\begin{pmatrix}\mathcal{E}_{+} & \mathcal{E}_{-}\end{pmatrix}^{T}=\mathcal{R}^{-1}\overrightarrow{\mathcal{E}_{i}}$.
Eq.\ (\ref{eq:dEdz'}) may be replaced by two scalar equations:
\begin{eqnarray}
\left[\partial_{z'}-\frac{2\beta_{i}}{\omega_{i}}\left(G\mp\left|\rho\right|\right)\partial_{\tau}\right]\mathcal{E}_{\pm} & = & -i\beta_{i}\left(G\mp\left|\rho\right|\right)\mathcal{E}_{\pm}+\label{eq:simpler scalar}\\
 & + & 2\frac{\beta_{i}}{\omega_{i}}\left[\partial_{\tau}\left(G\mp\left|\rho\right|\right)\right]\mathcal{E}_{\pm}+\nonumber \\
 & + & i\beta_{i}\frac{\dot{\Phi}}{\omega_{i}}\left(G\pm\left|\rho\right|\right)\mathcal{E}_{\mp}\quad,\nonumber
\end{eqnarray}
 where $\Phi$ is the argument of $\rho\equiv\left|\rho\right|\exp i\Phi$.
Equations \eqref{eq:simpler scalar} (or Eqs.\ (\ref{eq:dEdz'}))
describe a rich variety of phenomena in a time-dependent birefringent
medium. In particular, the second term on the rhs of Eq.\ \eqref{eq:simpler scalar},
is responsible for amplification/attenuation of the field amplitudes
$\mathcal{E}_{\pm}$  in a non-stationary anisotropic  medium with "non-rotating" birefringence axes.
The third term on the rhs of Eq.\ \eqref{eq:simpler scalar}
describes a non-adiabatic coupling between the $\mathcal{E}_{\pm}$
amplitudes in a medium with rotating birefringence, which also results in a change of the CP amplitudes (see, e.\,g.\
\cite{Pippard1994}). However, the cumulative amplitudes' change
due to these terms is small as long as the difference in propagation
time of the two field components through the medium is smaller than
the molecular alignment time scale ( $l\ll\Delta t_{al}c/(N \left|\Delta \alpha \right| \epsilon_0^{-1} )$), and as
long as the frequency of the birefringence axis rotation is slower than the optical frequency,
$\left|\dot{\Phi}\right|\ll\omega_{i}$. Under these conditions, which
prevail in recent experiments \cite{Korech2013,Korobenko2013}, we
may neglect these terms.

The remaining uncoupled equations have a simple solution, which we
combine and obtain the total CP field at location $z$ as a function
of the incoming field envelope $\overrightarrow{\mathcal{E}_{i}}\left(z=0\right)$:

\selectlanguage{english}%
{\small
\begin{eqnarray}
\mathbf{E}_{i}\left(z\right) & \approx & e^{i\omega_{i}\tau}\exp\left(i\beta_{i}Gz\right)\times\label{eq:slowly solved}\\
 &  & \left[\cos\left(\beta_{i}\left|\rho\right|z\right)I+i\sin\left(\beta_{i}\left|\rho\right|z\right)\begin{pmatrix}0 & e^{-i\Phi}\\
e^{i\Phi} & 0
\end{pmatrix}\right]\overrightarrow{\mathcal{E}_{i}}\left(0\right)\enskip.\nonumber
\end{eqnarray}
}\foreignlanguage{american}{This expression describes a Rabi-like
oscillation between the two CP components. We identify the $z$-dependent
phase modulation terms, especially the exponent containing $G$. The
rotational Doppler shift effect is described by the exponents of $\Phi$
in the off-diagonal elements of the matrix in the right hand side
of Eq.~(\ref{eq:slowly solved}). To validate our solution, we have
analyzed polarization and spectral transformations of a probe pulse
for various medium-preparation scenarios (see below) and found a good
agreement between the results based on the analytical Eq.\ (\ref{eq:slowly solved})
and those obtained by solving Eq.\ (\ref{eq:simpler scalar}) numerically
(finite difference time domain calculations). }

\selectlanguage{american}%
Next we discuss the effect of specific molecular excitation scenarios
on the probe pulse. We first examine the simplest model case in which
the molecular orientation is restricted to be perpendicular to the
laser propagation direction ($z$ axis). We assume that when the maximum
of the probe pulse arrives, the molecules are perfectly aligned and
have a normal distribution of angular velocity $\dot{\varphi}$ with
a non-zero average $\Omega$ and a standard deviation $\sigma$. It
can be shown that in this case $\left|\rho\right|=\exp\left(-2\tau^{2}\sigma^{2}\right)$,
$\Phi=2\Omega\tau$ and $G=0$. For a left-CP input field $\mathcal{E}_{L}\left(z=0\right)$,
Eq.\ (\ref{eq:slowly solved}) provides the following expression
for the generated right-CP field at the exit of the medium: %
\begin{comment}
pg 180 august 12
\end{comment}
\begin{eqnarray}
E{}_{R}\left(z\right) & \approx & i\sin\left(z\beta_{i}e^{-2\tau^{2}\sigma^{2}}\right)e^{i\left(\omega+2\Omega\right)\tau}\mathcal{E}_{L}\left(0\right)\quad.\label{eq:general ensemble}
\end{eqnarray}
The apparent $+2\Omega$ frequency shift in the generated field is
inverted when the input field has a right-CP handedness. Moreover,
when the angular velocity deviation $\sigma$ is zero (all the molecules
rotate in unison, and with the same speed), Eq.\ (\ref{eq:general ensemble}) shows a
simple Rabi-type oscillation and frequency shift that is consistent with
that of light propagating through a rotating waveplate \cite{Garetz1979,Tudor2001}.
Our simplified classical model of a "molecular waveplate"  corresponds to experiment \cite{Harris},
in which highly efficient single-sideband frequency conversion was indeed observed via coherent driving of a single rovibrational Raman transition
in molecular deuterium.

\begin{figure}
\includegraphics[width=3.3in]{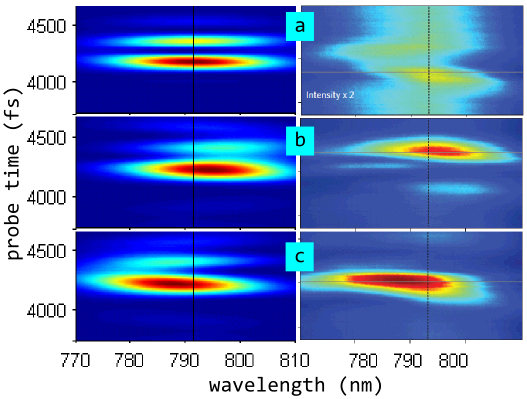}

\caption{\label{fig:Theoretical-spectrum-1}(Color online) Spectrum
of the output field of inverted circular polarization vs. pump-probe delay for $^{14}\textrm{N}_{2}$ molecules
brought to unidirectional rotation by the two-pulse "molecular propeller" scheme.
Color depicts the magnitude of the signal (arbitrary scaling).
Left column shows calculated spectrum, the right one presents experimental data from \cite{Korech2013} (with permission).
Panel (a) displays results for a control single-pulse excitation for which no chiral effect exists.
Panel (b) corresponds to molecules having the same sense of rotation as the electric vector of the input CP pulse,
panel (c) presents results for the opposite sense of molecular rotation. }

\end{figure}

We proceed now to a more sophisticated 3D model of molecules brought into fast
unidirectional rotation by a pair of delayed pump pulses according
to the {}`molecular propeller' scheme \cite{Fleischer2009,Kitano2009,benzene}.
To calculate the $\rho$ and $G$ time-dependent values, we used both
classical Monte Carlo simulation \cite{benzene} (for short pump-probe
delays) and full quantum mechanical calculation (when the delay becomes
comparable to the molecular rotational revival time). In the latter
case, we analyzed the rotational dynamics of quantum wave packets
and averaged over the thermal distribution of the initial rotational
states (similarly to \cite{Fleischer2009}).

Figure \ref{fig:Theoretical-spectrum-1} depicts the calculated frequency
content of the output pulse of inverted circular polarization for
different time delays between the CP probe and the last pump pulse (left panels), as compared to the recent experiment  \cite{Korech2013} (right hand side).
This example corresponds to nitrogen molecules of air at ambient conditions,
and the delays are chosen around the half-revival time of the nitrogen,
$T_{rev}=8.3\, ps$ \cite{Stapelfeldt2003}. The kick strength \cite{Leibscher2003}
of each of the pump pulses is $P=5$, close to the estimated experimental value. Figure  \ref{fig:Theoretical-spectrum-1}a
shows a spectrogram for a control case in which
only a single pump pulse is employed. The vertical line marks the central wavelength of the incident pulse. A considerable spectral broadening
is observed in the half-revival region (both in simulations and in the experiment) because of the phase modulation
effect, but no visible spectral shift is present. However, if unidirectional rotation is present(see the
middle and the lower panels of Fig.\ \ref{fig:Theoretical-spectrum-1}), the spectrograms
are shifted to the red or to the blue, depending on the relative handedness
of the probe field polarization and induced molecular rotation.
In our  simulation, the rotational Doppler shift reaches
the level of about 5 THz. For a $800\, nm$ probe pulse this translates
to the wavelength deviation of about $\sim10\, nm$, close to the
spectral width of pulses used in experiments \cite{Korech2013}. The predicted value of the shift and its direction are in a good agreement
with the reported experimental observations (see Figs. \ \ref{fig:Theoretical-spectrum-1}b,c).

As an additional and independent test of our theory, we consider another recent experiment  \cite{Korobenko2013},
in which unidirectionally rotating $\textrm{O}_{2}$ molecules were
produced using the {}`optical centrifuge' scheme. In that experiment,
molecules were optically spun at a continuously increasing rotational
frequency, up to $J\sim69\hbar$ after which the centrifuge was abruptly
switched off (only  odd values of $J$ are allowed for $^{16}\textrm{O}_{2}$ molecule because of its nuclear spin statistics). The optical properties of the molecular medium were
probed by a delayed CP  probe pulse during the centrifuge operation, and after its termination.

\begin{figure}
\includegraphics[width=3.4in]{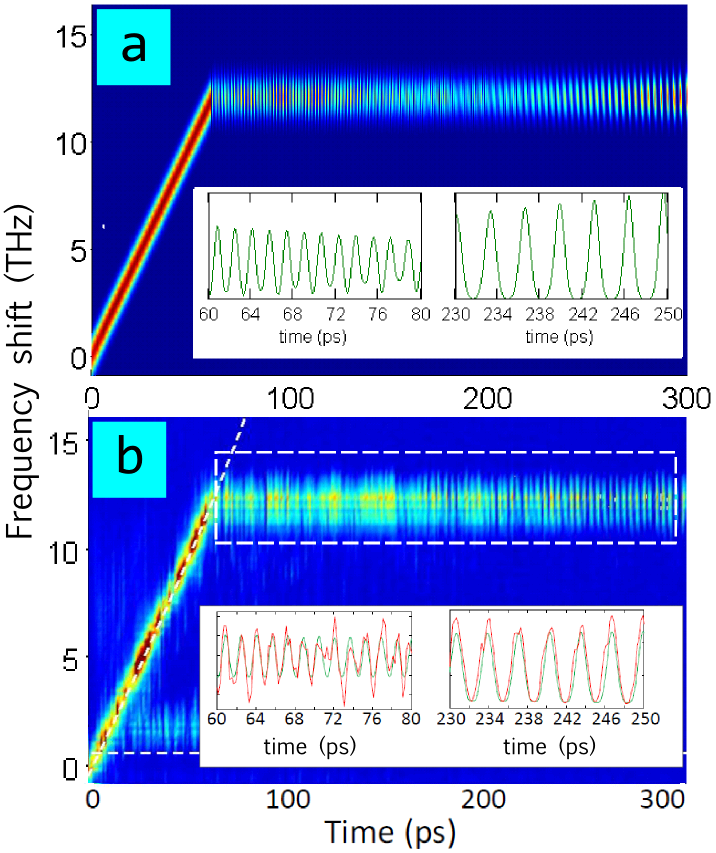}

\caption{\label{fig:optical centr}(Color online) (a) Calculated spectrum
of the output field of inverted circular polarization vs. probe delay for
$^{16}\textrm{O}_{2}$ molecules spun by the `optical centrifuge' to $J\sim69\hbar$.
(b) The corresponding experimental results (from \cite{Korobenko2013}, with permission).
 The insets show the signal intensity (a.u.) after the end of centrifugation. }
\end{figure}

To model this situation, we analyzed the time dependence of the functions $G$ and $\rho$ during the acceleration period (when the molecules are trapped by the centrifuge field)
both classically and quantum mechanically, with a good agreement between the two approaches.
The long-time evolution of these functions after the driving field was cut off and acceleration had stopped was
treated quantum mechanically, in order to account for quantum revivals
of the released rotational wave packets. Figure \ref{fig:optical centr}a
shows our results for the spectrum of the oppositely circularly polarized signal at the
output of the medium as a function of the probe delay with respect to the start of the centrifuge.
 The calculated signal shows  good qualitative and quantitative agreement with the results of  the reported experimental
observations \cite{Korobenko2013} (see Fig.\ \ref{fig:optical centr}b).
At the acceleration stage, it exhibits linearly growing frequency shift corresponding to twice the instantaneous rotation frequency of the centrifuge.
In the field-free regime (after the end of the centrifuge at $\sim 60\ ps$), the excited rotational wavepacket produces a revival oscillatory pattern of the signal strength, as shown in the insets of Fig.\ \ref{fig:optical centr}.
The revival period changes with time because of the centrifugal distortion effect in the molecular rotational  spectrum. The exact position of the revival peaks depends on the relative phases accumulated by different  rotational states
during the centrifugation and in the process of the release from the centrifuge \cite{Korobenko2013}.  We matched our results to the experiment by tuning the phases of  different  states of the final rotational wavepacket
to fit the position of a single, arbitrarily chosen peak of the signal experimentally recorded near $\sim300\ ps$ delay. As a result, a very good agreement between the calculated and measured revival signals has been achieved in the whole
post-centrifugation region (compare the insets of Figs.\ \ref{fig:optical centr}a and \ref{fig:optical centr}b).

\begin{figure}
\includegraphics[width=2.5in]{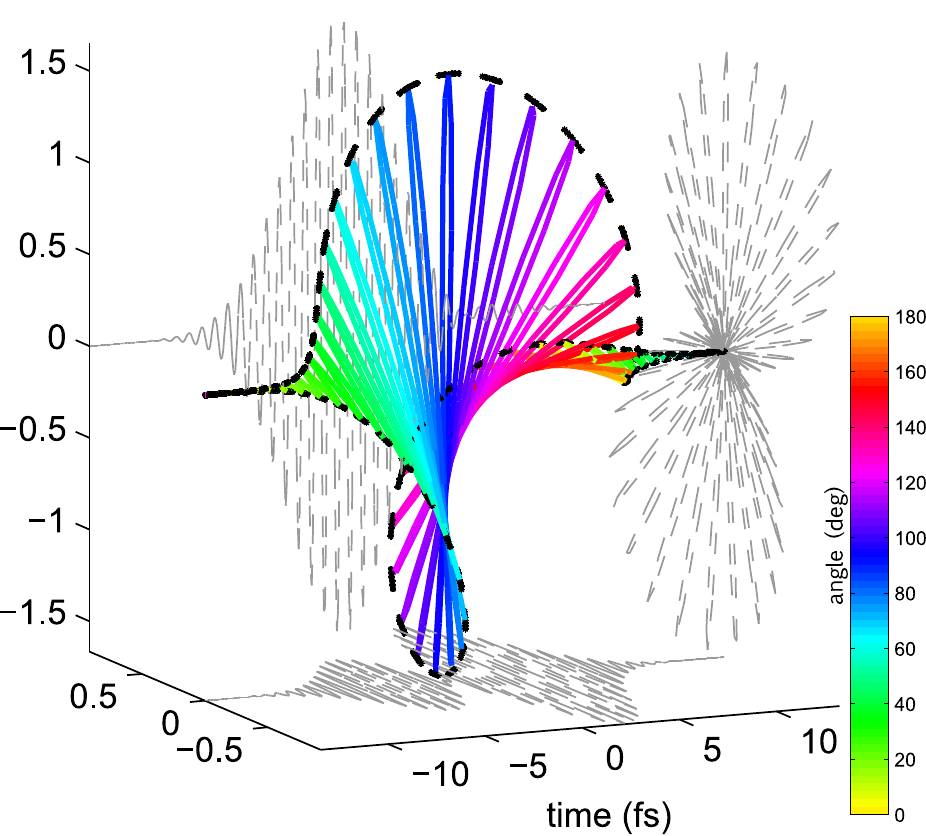}\foreignlanguage{english}{\caption{\selectlanguage{american}%
\label{fig:The-WARP-pulse}(Color online) The electric field of the WARP pulse
(arbitrary field units, projections in a lighter shade, color represents the instantaneous linear polarization angle) resulting from
the propagation of a linearly polarized probe through a gas of unidirectionally rotating nitrogen
molecules. The simulation was performed assuming a double-pulse "molecular propeller" excitation scheme
 at atmospheric conditions.\foreignlanguage{english}{ }\selectlanguage{english}
}
}
\end{figure}

Finally, we discuss an additional manifestation of the chirality transfer
from the molecular medium  to a probe light, when the input probe is linearly polarized. The latter may be regarded as a pair of circularly polarized pulses of opposite handedness. When the probe encounters a gas of
unidirectionally rotating molecules, each of these CP components partially
transforms into a {}`daughter' CP pulse of opposite chirality.
The two mirror CP daughter pulses appear in phase, they have opposite rotational Doppler frequency shifts
but equal amplitudes, thus they effectively combine into a linearly polarized field. However, the polarization direction, determined
by the relative phase between the two pulses, rotates with time at  twice the mean frequency of the molecular spinning, resulting in a wave of autonomously rotating
linear polarization (WARP) featured in Fig.\ \ref{fig:The-WARP-pulse}. This kind of polychromatic light
has been created in the past \cite{Emile1997} using a mechanically rotating waveplate. The mechanisms considered in our paper
allow for generating the WARPs with the polarization rotation frequency which is many orders of magnitude higher.

In conclusion, we presented a theoretical analysis of the optical properties
of a gas of unidirectionally rotating molecules, and showed that such a coherent collective
spinning  substantially shifts the spectrum of a CP light pulse passing
through the medium. The direction of the shift is determined by the relative handedness of the molecular rotation and the CP pulse, and the shift value is controlled by the rate of molecular rotation.
Our   treatment considers in a unified way two recent experiments on the  optics
of gases with unidirectionally spinning molecules \cite{Korech2013,Korobenko2013},  and provides a good qualitative and quantitative
description of the measurement results. We also discussed using the molecular-induced frequency shift to prepare a pulse
whose linear polarization continuously rotates at a THz-range speed.
This `twisted polarization' mode may become an interesting and useful addition to the gallery of femtosecond shaped pulses, especially in
view of  high and tunable rotation frequency, and relatively simple preparation method.
\begin{acknowledgments}
We appreciate useful discussions with Johannes Flo\ss, Erez Gershnabel, Robert J.\,Gordon,
Yuri Khodorkovsky, Omer Korech, Aleksey Korobenko, Alexander Milner, and Valery Milner.
This work was partly supported by the Israel Science Foundation (ISF grant no. 601/10)),
the Deutsche Forschungsgemeinschaft (DFG grant no. LE 2138/2-1), and the historic generosity
of the Harold Perlman Family.
\end{acknowledgments}
%\bibliography{C:/Users/uri2x/Documents/literature/fromRIS}
\bibliographystyle{apsrev4-1}
%\bibliography{"frequency shift paper.bbl"}
%\bibliography{fromRIS-ilya.bbl}

%\bibliography{C:/Users/uri2x/Documents/literature/fromRIS}

\begin{thebibliography}{30}%
\makeatletter
\providecommand \@ifxundefined [1]{%
 \@ifx{#1\undefined}
}%
\providecommand \@ifnum [1]{%
 \ifnum #1\expandafter \@firstoftwo
 \else \expandafter \@secondoftwo
 \fi
}%
\providecommand \@ifx [1]{%
 \ifx #1\expandafter \@firstoftwo
 \else \expandafter \@secondoftwo
 \fi
}%
\providecommand \natexlab [1]{#1}%
\providecommand \enquote  [1]{``#1''}%
\providecommand \bibnamefont  [1]{#1}%
\providecommand \bibfnamefont [1]{#1}%
\providecommand \citenamefont [1]{#1}%
\providecommand \href@noop [0]{\@secondoftwo}%
\providecommand \href [0]{\begingroup \@sanitize@url \@href}%
\providecommand \@href[1]{\@@startlink{#1}\@@href}%
\providecommand \@@href[1]{\endgroup#1\@@endlink}%
\providecommand \@sanitize@url [0]{\catcode `\\12\catcode `\$12\catcode
  `\&12\catcode `\#12\catcode `\^12\catcode `\_12\catcode `\%12\relax}%
\providecommand \@@startlink[1]{}%
\providecommand \@@endlink[0]{}%
\providecommand \url  [0]{\begingroup\@sanitize@url \@url }%
\providecommand \@url [1]{\endgroup\@href {#1}{\urlprefix }}%
\providecommand \urlprefix  [0]{URL }%
\providecommand \Eprint [0]{\href }%
\providecommand \doibase [0]{http://dx.doi.org/}%
\providecommand \selectlanguage [0]{\@gobble}%
\providecommand \bibinfo  [0]{\@secondoftwo}%
\providecommand \bibfield  [0]{\@secondoftwo}%
\providecommand \translation [1]{[#1]}%
\providecommand \BibitemOpen [0]{}%
\providecommand \bibitemStop [0]{}%
\providecommand \bibitemNoStop [0]{.\EOS\space}%
\providecommand \EOS [0]{\spacefactor3000\relax}%
\providecommand \BibitemShut  [1]{\csname bibitem#1\endcsname}%
\let\auto@bib@innerbib\@empty
%</preamble>
\bibitem [{\citenamefont {Garetz}(1981)}]{Garetz1981a}%
  \BibitemOpen
  \bibfield  {author} {\bibinfo {author} {\bibfnamefont {B.~A.}\ \bibnamefont
  {Garetz}},\ }\href
  {http://www.opticsinfobase.org/abstract.cfm?URI=josa-71-5-609} {\bibfield
  {journal} {\bibinfo  {journal} {J. Opt. Soc. Am.}\ }\textbf {\bibinfo
  {volume} {71}},\ \bibinfo {pages} {609} (\bibinfo {year} {1981})}\BibitemShut
  {NoStop}%
\bibitem [{\citenamefont {Bretenaker}\ and\ \citenamefont
  {Le~Floch}(1990)}]{Bretenaker1990}%
  \BibitemOpen
  \bibfield  {author} {\bibinfo {author} {\bibfnamefont {F.}~\bibnamefont
  {Bretenaker}}\ and\ \bibinfo {author} {\bibfnamefont {A.}~\bibnamefont
  {Le~Floch}},\ }\href {\doibase 10.1103/PhysRevLett.65.2316} {\bibfield
  {journal} {\bibinfo  {journal} {Phys. Rev. Lett.}\ }\textbf {\bibinfo
  {volume} {65}},\ \bibinfo {pages} {2316} (\bibinfo {year}
  {1990})}\BibitemShut {NoStop}%
\bibitem [{\citenamefont {Allen}\ \emph {et~al.}(1994)\citenamefont {Allen},
  \citenamefont {Babiker},\ and\ \citenamefont {Power}}]{Allen1994}%
  \BibitemOpen
  \bibfield  {author} {\bibinfo {author} {\bibfnamefont {L.}~\bibnamefont
  {Allen}}, \bibinfo {author} {\bibfnamefont {M.}~\bibnamefont {Babiker}}, \
  and\ \bibinfo {author} {\bibfnamefont {W.~L.}\ \bibnamefont {Power}},\ }\href
  {http://www.sciencedirect.com/science/article/B6TVF-46JGKT4-2G/2/e43b87cf4d32d0cbe15f880a1a4537b3}
  {\bibfield  {journal} {\bibinfo  {journal} {Opt. Commun.}\ }\textbf {\bibinfo
  {volume} {112}},\ \bibinfo {pages} {141} (\bibinfo {year}
  {1994})}\BibitemShut {NoStop}%
\bibitem [{\citenamefont {Bialynicki-Birula}\ and\ \citenamefont
  {Bialynicka-Birula}(1997)}]{Bialynicki-Birula1997}%
  \BibitemOpen
  \bibfield  {author} {\bibinfo {author} {\bibfnamefont {I.}~\bibnamefont
  {Bialynicki-Birula}}\ and\ \bibinfo {author} {\bibfnamefont {Z.}~\bibnamefont
  {Bialynicka-Birula}},\ }\href {\doibase 10.1103/PhysRevLett.78.2539}
  {\bibfield  {journal} {\bibinfo  {journal} {Phys. Rev. Lett.}\ }\textbf
  {\bibinfo {volume} {78}},\ \bibinfo {pages} {2539} (\bibinfo {year}
  {1997})}\BibitemShut {NoStop}%
\bibitem [{\citenamefont {Bialynicki-Birula}\ and\ \citenamefont
  {Bialynicka-Birula}(2012)}]{Bialynicki-Birula2012}%
  \BibitemOpen
  \bibfield  {author} {\bibinfo {author} {\bibfnamefont {I.}~\bibnamefont
  {Bialynicki-Birula}}\ and\ \bibinfo {author} {\bibfnamefont {Z.}~\bibnamefont
  {Bialynicka-Birula}},\ }\href@noop {} {\bibfield  {journal} {\bibinfo
  {journal} {The Angular Momentum of Light, edited by Andrews, D.L. and
  Babiker, M.}\ ,\ \bibinfo {pages} {162}} (\bibinfo {year}
  {2012})}\BibitemShut {NoStop}%
\bibitem [{\citenamefont {Cabannes}\ and\ \citenamefont
  {Rocard}(1929)}]{Cabannes1929}%
  \BibitemOpen
  \bibfield  {author} {\bibinfo {author} {\bibfnamefont {J.}~\bibnamefont
  {Cabannes}}\ and\ \bibinfo {author} {\bibfnamefont {Y.}~\bibnamefont
  {Rocard}},\ }\href {\doibase 10.1051/jphysrad:0192900100205200} {\bibfield
  {journal} {\bibinfo  {journal} {J. Phys. Radium}\ }\textbf {\bibinfo {volume}
  {10}},\ \bibinfo {pages} {52} (\bibinfo {year} {1929})}\BibitemShut {NoStop}%
\bibitem [{\citenamefont {Allen}(1966)}]{Allen1966}%
  \BibitemOpen
  \bibfield  {author} {\bibinfo {author} {\bibfnamefont {P.~J.}\ \bibnamefont
  {Allen}},\ }\href {\doibase 10.1119/1.1972585} {\bibfield  {journal}
  {\bibinfo  {journal} {American Journal of Physics}\ }\textbf {\bibinfo
  {volume} {34}},\ \bibinfo {pages} {1185} (\bibinfo {year}
  {1966})}\BibitemShut {NoStop}%
\bibitem [{\citenamefont {Garetz}\ and\ \citenamefont
  {Arnold}(1979)}]{Garetz1979}%
  \BibitemOpen
  \bibfield  {author} {\bibinfo {author} {\bibfnamefont {B.~A.}\ \bibnamefont
  {Garetz}}\ and\ \bibinfo {author} {\bibfnamefont {S.}~\bibnamefont
  {Arnold}},\ }\href
  {http://www.sciencedirect.com/science/article/pii/003040187990230X}
  {\bibfield  {journal} {\bibinfo  {journal} {Opt. Commun.}\ }\textbf {\bibinfo
  {volume} {31}},\ \bibinfo {pages} {1} (\bibinfo {year} {1979})}\BibitemShut
  {NoStop}%
\bibitem [{\citenamefont {Courtial}\ \emph {et~al.}(1998)\citenamefont
  {Courtial}, \citenamefont {Robertson}, \citenamefont {Dholakia},
  \citenamefont {Allen},\ and\ \citenamefont {Padgett}}]{Courtial1998}%
  \BibitemOpen
  \bibfield  {author} {\bibinfo {author} {\bibfnamefont {J.}~\bibnamefont
  {Courtial}}, \bibinfo {author} {\bibfnamefont {D.~A.}\ \bibnamefont
  {Robertson}}, \bibinfo {author} {\bibfnamefont {K.}~\bibnamefont {Dholakia}},
  \bibinfo {author} {\bibfnamefont {L.}~\bibnamefont {Allen}}, \ and\ \bibinfo
  {author} {\bibfnamefont {M.~J.}\ \bibnamefont {Padgett}},\ }\href
  {http://link.aps.org/doi/10.1103/PhysRevLett.81.4828} {\bibfield  {journal}
  {\bibinfo  {journal} {Phys. Rev. Lett.}\ }\textbf {\bibinfo {volume} {81}},\
  \bibinfo {pages} {4828} (\bibinfo {year} {1998})}\BibitemShut {NoStop}%
\bibitem [{\citenamefont {Lavery}\ \emph {et~al.}(2013)\citenamefont {Lavery},
  \citenamefont {Speirits}, \citenamefont {Barnett},\ and\ \citenamefont
  {Padgett}}]{Lavery2013}%
  \BibitemOpen
  \bibfield  {author} {\bibinfo {author} {\bibfnamefont {M.~P.~J.}\
  \bibnamefont {Lavery}}, \bibinfo {author} {\bibfnamefont {F.~C.}\
  \bibnamefont {Speirits}}, \bibinfo {author} {\bibfnamefont {S.~M.}\
  \bibnamefont {Barnett}}, \ and\ \bibinfo {author} {\bibfnamefont {M.~J.}\
  \bibnamefont {Padgett}},\ }\href@noop {} {\bibfield  {journal} {\bibinfo
  {journal} {Science}\ }\textbf {\bibinfo {volume} {341}},\ \bibinfo {pages}
  {537} (\bibinfo {year} {2013})}\BibitemShut {NoStop}%
\bibitem [{\citenamefont {Buhrer}\ \emph {et~al.}(1962)\citenamefont {Buhrer},
  \citenamefont {Baird},\ and\ \citenamefont {Conwell}}]{Buhrer1962}%
  \BibitemOpen
  \bibfield  {author} {\bibinfo {author} {\bibfnamefont {C.~F.}\ \bibnamefont
  {Buhrer}}, \bibinfo {author} {\bibfnamefont {D.}~\bibnamefont {Baird}}, \
  and\ \bibinfo {author} {\bibfnamefont {E.~M.}\ \bibnamefont {Conwell}},\
  }\href {http://link.aip.org/link/?APL/1/46/1} {\bibfield  {journal} {\bibinfo
   {journal} {Appl. Phys. Lett.}\ }\textbf {\bibinfo {volume} {1}},\ \bibinfo
  {pages} {46} (\bibinfo {year} {1962})}\BibitemShut {NoStop}%
\bibitem [{\citenamefont {Fleischer}\ \emph {et~al.}(2012)\citenamefont
  {Fleischer}, \citenamefont {Khodorkovsky}, \citenamefont {Gershnabel},
  \citenamefont {Prior},\ and\ \citenamefont {{I. Sh.
  Averbukh}}}]{Fleischer2012}%
  \BibitemOpen
  \bibfield  {author} {\bibinfo {author} {\bibfnamefont {S.}~\bibnamefont
  {Fleischer}}, \bibinfo {author} {\bibfnamefont {Y.}~\bibnamefont
  {Khodorkovsky}}, \bibinfo {author} {\bibfnamefont {E.}~\bibnamefont
  {Gershnabel}}, \bibinfo {author} {\bibfnamefont {Y.}~\bibnamefont {Prior}}, \
  and\ \bibinfo {author} {\bibnamefont {{I. Sh. Averbukh}}},\ }\href@noop {}
  {\bibfield  {journal} {\bibinfo  {journal} {Isr. J. Chem.}\ }\textbf
  {\bibinfo {volume} {52}},\ \bibinfo {pages} {414 } (\bibinfo {year}
  {2012})}\BibitemShut {NoStop}%
\bibitem [{\citenamefont {Ohshima}\ and\ \citenamefont
  {Hasegawa}(2010)}]{Ohshima2010}%
  \BibitemOpen
  \bibfield  {author} {\bibinfo {author} {\bibfnamefont {Y.}~\bibnamefont
  {Ohshima}}\ and\ \bibinfo {author} {\bibfnamefont {H.}~\bibnamefont
  {Hasegawa}},\ }\href {\doibase 10.1080/0144235X.2010.511769} {\bibfield
  {journal} {\bibinfo  {journal} {International Reviews in Physical Chemistry}\
  }\textbf {\bibinfo {volume} {29}},\ \bibinfo {pages} {619} (\bibinfo {year}
  {2010})}\BibitemShut {NoStop}%
\bibitem [{\citenamefont {Stapelfeldt}\ and\ \citenamefont
  {Seideman}(2003)}]{Stapelfeldt2003}%
  \BibitemOpen
  \bibfield  {author} {\bibinfo {author} {\bibfnamefont {H.}~\bibnamefont
  {Stapelfeldt}}\ and\ \bibinfo {author} {\bibfnamefont {T.}~\bibnamefont
  {Seideman}},\ }\href {http://link.aps.org/doi/10.1103/RevModPhys.75.543}
  {\bibfield  {journal} {\bibinfo  {journal} {Rev. Mod. Phys.}\ }\textbf
  {\bibinfo {volume} {75}},\ \bibinfo {pages} {543} (\bibinfo {year}
  {2003})}\BibitemShut {NoStop}%
\bibitem [{\citenamefont {Karczmarek}\ \emph {et~al.}(1999)\citenamefont
  {Karczmarek}, \citenamefont {Wright}, \citenamefont {Corkum},\ and\
  \citenamefont {Ivanov}}]{Karczmarek1999}%
  \BibitemOpen
  \bibfield  {author} {\bibinfo {author} {\bibfnamefont {J.}~\bibnamefont
  {Karczmarek}}, \bibinfo {author} {\bibfnamefont {J.}~\bibnamefont {Wright}},
  \bibinfo {author} {\bibfnamefont {P.}~\bibnamefont {Corkum}}, \ and\ \bibinfo
  {author} {\bibfnamefont {M.}~\bibnamefont {Ivanov}},\ }\href {\doibase
  10.1103/PhysRevLett.82.3420} {\bibfield  {journal} {\bibinfo  {journal}
  {Phys. Rev. Lett.}\ }\textbf {\bibinfo {volume} {82}},\ \bibinfo {pages}
  {3420} (\bibinfo {year} {1999})}\BibitemShut {NoStop}%
\bibitem [{\citenamefont {Villeneuve}\ \emph {et~al.}(2000)\citenamefont
  {Villeneuve}, \citenamefont {Aseyev}, \citenamefont {Dietrich}, \citenamefont
  {Spanner}, \citenamefont {Ivanov},\ and\ \citenamefont
  {Corkum}}]{Villeneuve2000}%
  \BibitemOpen
  \bibfield  {author} {\bibinfo {author} {\bibfnamefont {D.~M.}\ \bibnamefont
  {Villeneuve}}, \bibinfo {author} {\bibfnamefont {S.~A.}\ \bibnamefont
  {Aseyev}}, \bibinfo {author} {\bibfnamefont {P.}~\bibnamefont {Dietrich}},
  \bibinfo {author} {\bibfnamefont {M.}~\bibnamefont {Spanner}}, \bibinfo
  {author} {\bibfnamefont {M.~Y.}\ \bibnamefont {Ivanov}}, \ and\ \bibinfo
  {author} {\bibfnamefont {P.~B.}\ \bibnamefont {Corkum}},\ }\href
  {http://link.aps.org/doi/10.1103/PhysRevLett.85.542} {\bibfield  {journal}
  {\bibinfo  {journal} {Phys. Rev. Lett.}\ }\textbf {\bibinfo {volume} {85}},\
  \bibinfo {pages} {542} (\bibinfo {year} {2000})}\BibitemShut {NoStop}%
\bibitem [{\citenamefont {Yuan}\ \emph {et~al.}(2011)\citenamefont {Yuan},
  \citenamefont {Teitelbaum}, \citenamefont {Robinson},\ and\ \citenamefont
  {Mullin}}]{Yuan2011}%
  \BibitemOpen
  \bibfield  {author} {\bibinfo {author} {\bibfnamefont {L.}~\bibnamefont
  {Yuan}}, \bibinfo {author} {\bibfnamefont {S.~W.}\ \bibnamefont
  {Teitelbaum}}, \bibinfo {author} {\bibfnamefont {A.}~\bibnamefont
  {Robinson}}, \ and\ \bibinfo {author} {\bibfnamefont {A.~S.}\ \bibnamefont
  {Mullin}},\ }\href {\doibase 10.1073/pnas.1018669108} {\bibfield  {journal}
  {\bibinfo  {journal} {Proc. Natl. Acad. Sci. USA}\ }\textbf {\bibinfo
  {volume} {108}},\ \bibinfo {pages} {6872} (\bibinfo {year}
  {2011})}\BibitemShut {NoStop}%
\bibitem [{\citenamefont {Fleischer}\ \emph {et~al.}(2009)\citenamefont
  {Fleischer}, \citenamefont {Khodorkovsky}, \citenamefont {Prior},\ and\
  \citenamefont {{I. Sh. Averbukh}}}]{Fleischer2009}%
  \BibitemOpen
  \bibfield  {author} {\bibinfo {author} {\bibfnamefont {S.}~\bibnamefont
  {Fleischer}}, \bibinfo {author} {\bibfnamefont {Y.}~\bibnamefont
  {Khodorkovsky}}, \bibinfo {author} {\bibfnamefont {Y.}~\bibnamefont {Prior}},
  \ and\ \bibinfo {author} {\bibnamefont {{I. Sh. Averbukh}}},\ }\href
  {\doibase 10.1088/1367-2630/11/10/105039} {\bibfield  {journal} {\bibinfo
  {journal} {New J. Phys.}\ }\textbf {\bibinfo {volume} {11}},\ \bibinfo
  {pages} {15} (\bibinfo {year} {2009})}\BibitemShut {NoStop}%
\bibitem [{\citenamefont {Kitano}\ \emph {et~al.}(2009)\citenamefont {Kitano},
  \citenamefont {Hasegawa},\ and\ \citenamefont {Ohshima}}]{Kitano2009}%
  \BibitemOpen
  \bibfield  {author} {\bibinfo {author} {\bibfnamefont {K.}~\bibnamefont
  {Kitano}}, \bibinfo {author} {\bibfnamefont {H.}~\bibnamefont {Hasegawa}}, \
  and\ \bibinfo {author} {\bibfnamefont {Y.}~\bibnamefont {Ohshima}},\ }\href
  {http://link.aps.org/doi/10.1103/PhysRevLett.103.223002} {\bibfield
  {journal} {\bibinfo  {journal} {Phys. Rev. Lett.}\ }\textbf {\bibinfo
  {volume} {103}},\ \bibinfo {pages} {223002} (\bibinfo {year}
  {2009})}\BibitemShut {NoStop}%
\bibitem [{\citenamefont {Khodorkovsky}\ \emph {et~al.}(2011)\citenamefont
  {Khodorkovsky}, \citenamefont {Kitano}, \citenamefont {Hasegawa},
  \citenamefont {Ohshima},\ and\ \citenamefont {{I. Sh. Averbukh}}}]{benzene}%
  \BibitemOpen
  \bibfield  {author} {\bibinfo {author} {\bibfnamefont {Y.}~\bibnamefont
  {Khodorkovsky}}, \bibinfo {author} {\bibfnamefont {K.}~\bibnamefont
  {Kitano}}, \bibinfo {author} {\bibfnamefont {H.}~\bibnamefont {Hasegawa}},
  \bibinfo {author} {\bibfnamefont {Y.}~\bibnamefont {Ohshima}}, \ and\
  \bibinfo {author} {\bibnamefont {{I. Sh. Averbukh}}},\ }\href@noop {}
  {\bibfield  {journal} {\bibinfo  {journal} {Physical Review A}\ }\textbf
  {\bibinfo {volume} {83}},\ \bibinfo {pages} {023423} (\bibinfo {year}
  {2011})}\BibitemShut {NoStop}%
\bibitem [{\citenamefont {Zhdanovich}\ \emph {et~al.}(2011)\citenamefont
  {Zhdanovich}, \citenamefont {Milner}, \citenamefont {Bloomquist},
  \citenamefont {Flo\ss}, \citenamefont {{I. Sh. Averbukh}}, \citenamefont
  {Hepburn},\ and\ \citenamefont {Milner}}]{Zhdanovich2011}%
  \BibitemOpen
  \bibfield  {author} {\bibinfo {author} {\bibfnamefont {S.}~\bibnamefont
  {Zhdanovich}}, \bibinfo {author} {\bibfnamefont {A.~A.}\ \bibnamefont
  {Milner}}, \bibinfo {author} {\bibfnamefont {C.}~\bibnamefont {Bloomquist}},
  \bibinfo {author} {\bibfnamefont {J.}~\bibnamefont {Flo\ss}}, \bibinfo
  {author} {\bibnamefont {{I. Sh. Averbukh}}}, \bibinfo {author} {\bibfnamefont
  {J.~W.}\ \bibnamefont {Hepburn}}, \ and\ \bibinfo {author} {\bibfnamefont
  {V.}~\bibnamefont {Milner}},\ }\href
  {http://link.aps.org/doi/10.1103/PhysRevLett.107.243004} {\bibfield
  {journal} {\bibinfo  {journal} {Phys. Rev. Lett.}\ }\textbf {\bibinfo
  {volume} {107}},\ \bibinfo {pages} {243004} (\bibinfo {year}
  {2011})}\BibitemShut {NoStop}%
\bibitem [{\citenamefont {Bloomquist}\ \emph {et~al.}(2012)\citenamefont
  {Bloomquist}, \citenamefont {Zhdanovich}, \citenamefont {Milner},\ and\
  \citenamefont {Milner}}]{Bloomquist2012}%
  \BibitemOpen
  \bibfield  {author} {\bibinfo {author} {\bibfnamefont {C.}~\bibnamefont
  {Bloomquist}}, \bibinfo {author} {\bibfnamefont {S.}~\bibnamefont
  {Zhdanovich}}, \bibinfo {author} {\bibfnamefont {A.}~\bibnamefont {Milner}},
  \ and\ \bibinfo {author} {\bibfnamefont {V.}~\bibnamefont {Milner}},\
  }\href@noop {} {\bibfield  {journal} {\bibinfo  {journal} {Phys. Rev. A}\
  }\textbf {\bibinfo {volume} {86}},\ \bibinfo {pages} {063413} (\bibinfo
  {year} {2012})}\BibitemShut {NoStop}%
\bibitem [{\citenamefont {Floss}\ and\ \citenamefont {{I. Sh.
  Averbukh}}(2012)}]{Floss2012}%
  \BibitemOpen
  \bibfield  {author} {\bibinfo {author} {\bibfnamefont {J.}~\bibnamefont
  {Floss}}\ and\ \bibinfo {author} {\bibnamefont {{I. Sh. Averbukh}}},\
  }\href@noop {} {\bibfield  {journal} {\bibinfo  {journal} {Phys. Rev. A}\
  }\textbf {\bibinfo {volume} {86}},\ \bibinfo {pages} {063414} (\bibinfo
  {year} {2012})}\BibitemShut {NoStop}%
\bibitem [{\citenamefont {Korech}\ \emph {et~al.}(2013)\citenamefont {Korech},
  \citenamefont {Steinitz}, \citenamefont {Gordon}, \citenamefont {{I. Sh.
  Averbukh}},\ and\ \citenamefont {Prior}}]{Korech2013}%
  \BibitemOpen
  \bibfield  {author} {\bibinfo {author} {\bibfnamefont {O.}~\bibnamefont
  {Korech}}, \bibinfo {author} {\bibfnamefont {U.}~\bibnamefont {Steinitz}},
  \bibinfo {author} {\bibfnamefont {R.~J.}\ \bibnamefont {Gordon}}, \bibinfo
  {author} {\bibnamefont {{I. Sh. Averbukh}}}, \ and\ \bibinfo {author}
  {\bibfnamefont {Y.}~\bibnamefont {Prior}},\ }\href@noop {} {\bibfield
  {journal} {\bibinfo  {journal} {Nature Photonics, DOI:
  10.1038/nphoton.2013.189; arXiv preprint arXiv:1303.6758}\ } (\bibinfo {year}
  {2013})}\BibitemShut {NoStop}%
\bibitem [{\citenamefont {Korobenko}\ \emph {et~al.}(2013)\citenamefont
  {Korobenko}, \citenamefont {Milner},\ and\ \citenamefont
  {Milner}}]{Korobenko2013}%
  \BibitemOpen
  \bibfield  {author} {\bibinfo {author} {\bibfnamefont {A.}~\bibnamefont
  {Korobenko}}, \bibinfo {author} {\bibfnamefont {A.~A.}\ \bibnamefont
  {Milner}}, \ and\ \bibinfo {author} {\bibfnamefont {V.}~\bibnamefont
  {Milner}},\ }\href@noop {} {\bibfield  {journal} {\bibinfo  {journal} {arXiv
  preprint arXiv:1304.0438}\ } (\bibinfo {year} {2013})}\BibitemShut {NoStop}%
\bibitem [{\citenamefont {Pippard}(1994)}]{Pippard1994}%
  \BibitemOpen
  \bibfield  {author} {\bibinfo {author} {\bibfnamefont {A.~B.}\ \bibnamefont
  {Pippard}},\ }\href {http://stacks.iop.org/0143-0807/15/i=2/a=007} {\bibfield
   {journal} {\bibinfo  {journal} {European Journal of Physics}\ }\textbf
  {\bibinfo {volume} {15}},\ \bibinfo {pages} {79} (\bibinfo {year}
  {1994})}\BibitemShut {NoStop}%
\bibitem [{\citenamefont {Tudor}(2001)}]{Tudor2001}%
  \BibitemOpen
  \bibfield  {author} {\bibinfo {author} {\bibfnamefont {T.}~\bibnamefont
  {Tudor}},\ }\href {\doibase 10.1364/JOSAA.18.000926} {\bibfield  {journal}
  {\bibinfo  {journal} {J. Opt. Soc. Am. A}\ }\textbf {\bibinfo {volume}
  {18}},\ \bibinfo {pages} {926} (\bibinfo {year} {2001})}\BibitemShut
  {NoStop}%
\bibitem [{\citenamefont {Sokolov}\ \emph {et~al.}(2001)\citenamefont
  {Sokolov}, \citenamefont {Sharpe}, \citenamefont {Shverdin}, \citenamefont
  {Walker}, \citenamefont {Yavuz}, \citenamefont {Yin},\ and\ \citenamefont
  {Harris}}]{Harris}%
  \BibitemOpen
  \bibfield  {author} {\bibinfo {author} {\bibfnamefont {A.~V.}\ \bibnamefont
  {Sokolov}}, \bibinfo {author} {\bibfnamefont {S.~J.}\ \bibnamefont {Sharpe}},
  \bibinfo {author} {\bibfnamefont {M.}~\bibnamefont {Shverdin}}, \bibinfo
  {author} {\bibfnamefont {D.~R.}\ \bibnamefont {Walker}}, \bibinfo {author}
  {\bibfnamefont {D.~D.}\ \bibnamefont {Yavuz}}, \bibinfo {author}
  {\bibfnamefont {G.~Y.}\ \bibnamefont {Yin}}, \ and\ \bibinfo {author}
  {\bibfnamefont {S.~E.}\ \bibnamefont {Harris}},\ }\href {\doibase
  10.1364/OL.26.000728} {\bibfield  {journal} {\bibinfo  {journal} {Opt.
  Lett.}\ }\textbf {\bibinfo {volume} {26}},\ \bibinfo {pages} {728} (\bibinfo
  {year} {2001})}\BibitemShut {NoStop}%
\bibitem [{\citenamefont {Leibscher}\ \emph {et~al.}(2003)\citenamefont
  {Leibscher}, \citenamefont {{I. Sh. Averbukh}},\ and\ \citenamefont
  {Rabitz}}]{Leibscher2003}%
  \BibitemOpen
  \bibfield  {author} {\bibinfo {author} {\bibfnamefont {M.}~\bibnamefont
  {Leibscher}}, \bibinfo {author} {\bibnamefont {{I. Sh. Averbukh}}}, \ and\
  \bibinfo {author} {\bibfnamefont {H.}~\bibnamefont {Rabitz}},\ }\href
  {\doibase 10.1103/PhysRevLett.90.213001} {\bibfield  {journal} {\bibinfo
  {journal} {Phys. Rev. Lett.}\ }\textbf {\bibinfo {volume} {90}},\ \bibinfo
  {pages} {213001} (\bibinfo {year} {2003})}\BibitemShut {NoStop}%
\bibitem [{\citenamefont {Emile}\ \emph {et~al.}(1997)\citenamefont {Emile},
  \citenamefont {Bretenaker},\ and\ \citenamefont {Le~Floch}}]{Emile1997}%
  \BibitemOpen
  \bibfield  {author} {\bibinfo {author} {\bibfnamefont {O.}~\bibnamefont
  {Emile}}, \bibinfo {author} {\bibfnamefont {F.}~\bibnamefont {Bretenaker}}, \
  and\ \bibinfo {author} {\bibfnamefont {A.}~\bibnamefont {Le~Floch}},\ }\href
  {\doibase 10.1142/S0217984997000293} {\bibfield  {journal} {\bibinfo
  {journal} {Mod. Phys. Lett. B}\ }\textbf {\bibinfo {volume} {11}},\ \bibinfo
  {pages} {219} (\bibinfo {year} {1997})}\BibitemShut {NoStop}%
\end{thebibliography}

\end{document}